\documentclass[applemac]{aa}
\usepackage{inputenc}
\usepackage{natbib}
\usepackage{txfonts}
\usepackage{graphicx}
\usepackage{color}
\bibpunct{(}{)}{;}{a}{}{,}

\def\arrow{\mathop{\rightarrow}\limits}
\def\htwo{\rm H_2}
\def\hthree{\rm H_3^+}
\def\htwod{\rm H_2D^+}
\def\dtwoh{\rm D_2H^+}

\begin{document}

\title{Species-to-species rate coefficients for the $\rm H_3^+ + H_2$ reacting system}
\author{O. Sipil\"a \inst{1},
		J. Harju\inst{1,2},
		\and{P. Caselli\inst{1}}
}
\institute{Max-Planck-Institute for Extraterrestrial Physics (MPE), Giessenbachstr. 1, 85748 Garching, Germany \\
e-mail: \texttt{osipila@mpe.mpg.de}
\and{Department of Physics, P.O. Box 64, 00014 University of Helsinki, Finland}
}

\date{Received / Accepted}

\authorrunning{O. Sipil\"a et al.}

\abstract
{}
{We study whether rotational excitation can make a large difference to chemical models of the abundances of the $\rm H_3^+$ isotopologs, including spin states, in physical conditions corresponding to starless cores and protostellar envelopes.}
{We developed a new rate coefficient set for the chemistry of the $\rm H_3^+$ isotopologs, allowing for rotational excitation, using the state-to-state rate coefficients published previously by Hugo et al. These new so-called species-to-species rate coefficients are compared with previously-used ground state-to-species rate coefficients by calculating chemical evolution in variable physical conditions using a pseudo-time-dependent chemical code.}
{We find that the new species-to-species model produces different results than the ground state-to-species model at high density and toward increasing temperatures ($T > 10$\,K). The most prominent difference is that the species-to-species model predicts a lower $\rm H_3^+$ deuteration degree at high density owing to an increase of the rate coefficients of endothermic reactions that tend to decrease deuteration. For example at 20\,K, the ground state-to-species model overestimates the abundance of $\rm H_2D^+$ by a factor of about two, while the abundance of $\rm D_3^+$ can differ by up to an order of magnitude between the models. The spin-state abundance ratios of the various $\rm H_3^+$ isotopologs are also affected, and the new model better reproduces recent observations of the abundances of ortho and para $\rm H_2D^+$ and $\rm D_2H^+$. The main caveat is that the applicability regime of the new rate coefficients depends on the critical densities of the various rotational transitions which vary with the abundances of the species and the temperature in dense clouds.}
{The difference in the abundances of the $\rm H_3^+$ isotopologs predicted by the species-to-species and ground state-to-species models is negligible at 10\,K corresponding to physical conditions in starless cores, but inclusion of the excited states is very important in studies of deuteration at higher temperatures, for example in protostellar envelopes. The species-to-species rate coefficients provide a more realistic approach to the chemistry of the $\rm H_3^+$ isotopologs than the ground state-to-species rate coefficients do, and so the former should be adopted in chemical models describing the chemistry of the $\rm H_3^+ + H_2$ reacting system.}


\maketitle

\section{Introduction}

The $\rm H_3^+$ ion is a universal proton donor and has therefore a central role in ion-molecule chemistry \citep{Herbst73}. In cold interstellar clouds, where the abundance of CO is reduced owing to its accretion onto dust, $\rm H_3^+$ becomes the most abundant cation and the principal distributor of deuterium from HD to other species (\citealt{Dalgarno84}; \citealt{Roberts03}) in reactions such as
\begin{equation}\label{h3+hd}
\rm H_3^+ + HD \rightleftharpoons H_2D^+ + H_2 \, ,
\end{equation}
which is exothermic by 232\,K when the reactants and products lie in their ground (para) states. The $\rm H_3^+$ ion also contributes strongly to the ortho-para conversion of $\rm H_2$ in these regions (\citealt{Flower06a}; \citealt{Pagani09}). (In what follows, the ortho and para states of each species are simply referred to as o and p, respectively.) Spin states play an important role in the development of deuterium chemistry, because the high energy of the rotational ground state of $\rm oH_2$ compared to that of $\rm pH_2$ ($\sim 170$\,K) can cause reaction~(\ref{h3+hd}) to proceed in the backward direction even at low temperatures, transferring deuterium back to HD.

The $\rm H_3^+$ isotopologs are among the most important tracers of the high density and low temperature regions of pre-stellar cores, where species heavier than He may be highly depleted because of freeze-out \citep{Caselli03, WFP04, Friesen14}. Also, a good understanding of spin state chemistry is fundamental because the spin ratios (in particular of $\rm H_2D^+$) and the linked deuteration fraction measured in, e.g., $\rm N_2H^+$, have been shown to be sensitive to the chemical age of dense clouds \citep{Brunken14, Kong15, Pagani11}. An accurate derivation of the chemical age of a cloud requires accurate chemical codes.

Investigation of chemical reaction dynamics is being pursued through quantum mechanical scattering calculations and state-to-state
experiments \citep[e.g.,][]{Teslja06, Zhang16}. Besides providing microscopic characterization of collisions between molecules, some of the results of these endeavors are directly applicable to astrophysical environments.  One such work is the study of the $\rm H_3^+ + H_2$ isotopic system by \cite{Hugo09}. The reaction between the trihydrogen cation, $\rm H_3^+$, and the hydrogen molecule, $\rm H_2$, is of fundamental importance for the physics and chemistry of interstellar clouds.

\cite{Hugo09} derived state-to-state thermal rate coefficients for inelastic and reactive collisions between all the isotopic variants of $\rm H_3^+$ and $\rm H_2$ and their different spin modifications. They compiled a table of ground state-to-species rate coefficients for reactive collisions, where it is assumed that all reacting isotopologs of $\rm H_3^+$ are in their ground rotational states. These data have been used in many studies of the chemistry of the $\rm H_3^+ + H_2$ reacting system \citep[e.g.,][]{Pagani09,Sipila10,Albertsson14b,LeGal14,Furuya15,Majumdar17}. The state-to-state rate coefficients can also be used to estimate species-to-species rate coefficients, where possible rotational excitation of the reactant species is explicitly taken into account. The assumption that the reactant species lie in their ground states is likely to be valid in the dense starless cores of molecular clouds with average densities of up to $n({\rm H_2}) \sim 10^5$ cm$^{-3}$, but in pre-stellar and star-forming cores with central densities exceeding $10^6$ cm$^{-3}$, some of the lowest rotationally excited levels of $\rm H_3^+$, $\rm H_2D^+$, $\rm D_2H^+$, and $\rm D_3^+$ should be excited and contribute to the total species-to-species reaction rates. Besides the construction of rate coefficients for chemical reactions, the state-to-state coefficients given by \citet{Hugo09} are needed for calculating the populations of the rotationally excited levels of $\rm H_2D^+$ and other isotopologs in connection with radiative transfer calculations.

A discrepancy between the predictions of chemical models utilizing either the ground state-to-species or species-to-species rate coefficients was recently suggested to arise in the protostellar core IRAS 16293-2422, where the model using the ground state-to-species rate coefficients seems to overpredict the $\rm oH_2D^+$ abundance \citep{Harju17b}. In the present paper, we discuss a chemical model which makes an effort to correct this effect by calculating the species-to-species rate coefficients from the state-to-state coefficients of \cite{Hugo09}, considering the density of the gas where the model is to be applied. This is done by estimating the populations of excited rotational states of $\rm H_3^+$, $\rm H_2D^+$, $\rm D_2H^+$, and $\rm D_3^+$ at different densities. For $\rm H_2D^+$ and $\rm D_2H^+$, the populations are obtained through radiative transfer modeling. For the symmetric ions $\rm H_3^+$ and $\rm D_3^+$ which have no electric dipole moments, the populations are assumed to obey the Boltzmann distribution.

The paper is organized as follows. We describe our models in Sect.\,\ref{s:models}. In Sect.\,\ref{s:results} we present our results, which are further discussed in Sect.\,\ref{s:discussion}. We give our conclusions in Sect\,\ref{s:conclusions}. Appendices~\ref{a:coeffderiv}~and~\ref{a:tables} contain additional discussion on critical densities, and rate coefficient tables.

\section{Model description}\label{s:models}

\subsection{Chemical model}\label{ss:chemmodel}

In this paper we aim to compare the abundances of the spin states of the $\rm H_3^+$ isotopologs as calculated with a model using either newly-calculated species-to-species rate coefficients or the previously-used ground state-to-species rate coefficients. We employ the gas-grain chemical model presented in detail in \citet{Sipila15a,Sipila15b}. Unless otherwise noted, we use the same physical parameters and initial abundances as given in Tables~1~and~3 in \citet{Sipila15a}. Here we use the KIDA gas-phase network \citep{Wakelam15} as the basis upon which the deuterium and spin-state chemistry is added according to the procedures discussed in detail in \citet{Sipila15a,Sipila15b}.

\subsection{New fits to the ground state-to-species rate coefficients}

\citet{Hugo09} calculated the state-to-state rate coefficients for the $\rm H_3^+ + H_2$ reacting system in the temperature range 5-50\,K. The data shown in their Table~VIII assumes that the reactants are in their rotational ground states. These so-called ground state-to-species rate coefficients were produced by fitting a two-parameter curve of the form $\alpha \exp(-\gamma/T)$ in the temperature range 5-20\,K\footnote{The fit was extended to 50\,K if the rates were lower than $10^{-17} \, \rm cm^3 \, s^{-1}$.} to data obtained from the state-to-state calculations. Later, \citet{Pagani13} have extended the fit to the range 5-50\,K for all included reactions using the modified Arrhenius rate law $k = \alpha \, (T/300)^\beta \, \exp(-\gamma/T)$.

Because the species-to-species rate coefficients (see Sect.\,\ref{ss:spectospeccoeff}) pertain to the temperature range 5-50\,K, for consistency we made a new fit to the ground state-to-species rates calculated by \citet{Hugo09} using the modified Arrhenius rate law like \citet{Pagani13} did. The resulting rate coefficients are given in Table~\ref{tabb1}. Unlike \citet{Pagani13}, we did not correct any of the fits by hand which leads to different values for the rate coefficients in some cases. We point out that there is a typo in Table~VIII in \citet{Hugo09}. The reaction $\rm pD_2H^+ + oH_2 \longrightarrow pH_3^+ + pD_2$ is erroneously marked as forbidden, while it is in fact allowed by spin selection rules (as also confirmed by the Hugo et al. state-so-state data). However, this reaction is insignificant as its rate coefficient is of the order of $10^{-18}\,\rm cm^3\,s^{-1}$ at 10\,K.

\begin{figure}
\centering
\includegraphics[width=1.0\columnwidth]{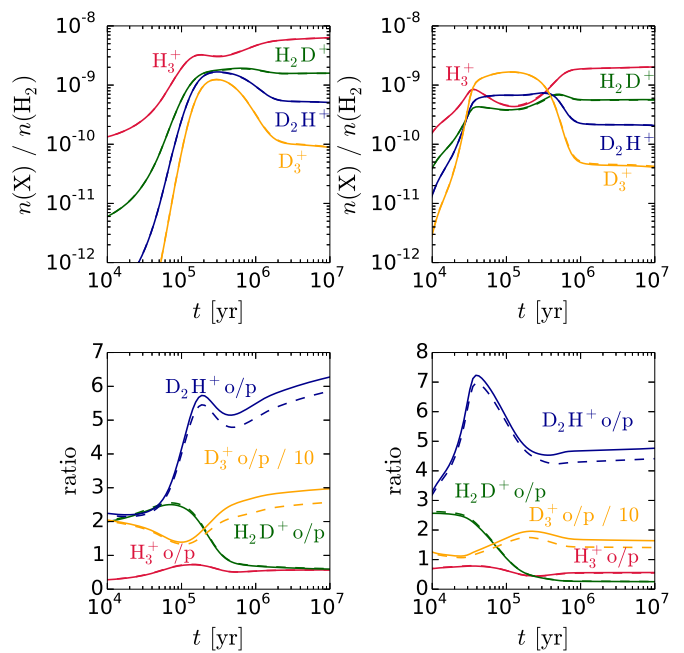}
\caption{Abundances ({\sl upper row}) and abundance ratios ({\sl lower row}) of the various $\rm H_3^+$ isotopologs as functions of time. The left-hand panels correspond to $n({\rm H_2}) = 10^5\,\rm cm^{-3}$, while the right-hand panels correspond to $n({\rm H_2}) = 10^6\,\rm cm^{-3}$. The temperature is set to $T_{\rm gas} = T_{\rm dust} = 10\,\rm K$. Solid lines represent calculations using the \citet{Hugo09} ground state-to-species rate coefficients, while dashed lines represent calculations using our new fit to the same coefficients (see text).
}
\label{fig:gstscomparison}
\end{figure}

Figure~\ref{fig:gstscomparison} shows a comparison of the abundances of the spin states of the $\rm H_3^+$ isotopologs calculated with the ground state-to-species rate coefficient fit from Table~VIII in \citet{Hugo09}, and our new fit. Evidently the total abundances of the $\rm H_3^+$ isotopologs are unaffected by the choice of rate coefficients, while the spin-state abundance ratios present slight deviations for the doubly and triply deuterated species (the variations for $\rm H_3^+$ and $\rm H_2D^+$ are not clearly visible in the plot). Calculations at different temperatures yield the same conclusions.

\begin{table*}
\caption{Critical densities ($n_{\rm c}$) of the excited rotational levels of $\rm H_2D^+$ and $\rm D_2H^+$, in order of increasing energy \citep{Hugo09}, used in the models presented in this paper. The corresponding rotational level ($J_{K_aK_c}$) is shown in parentheses after each value of critical density. The assumed fractional abundances are $X(\rm oH_2D^+) = 10^{-10}$, $X(\rm pH_2D^+) = 10^{-9}$, $X(\rm oD_2H^+) = 10^{-10}$, $X(\rm pD_2H^+) = 10^{-11}$, while the temperature is set to $T = 10$\,K.}

\centering
\begin{tabular}{c c c c c c c c}
\hline \hline 
Species & $n_{\rm c}\,[\rm cm^{-3}$] & Species & $n_{\rm c}\,[\rm cm^{-3}$] & Species & $n_{\rm c}\,[\rm cm^{-3}$] & Species & $n_{\rm c}\,[\rm cm^{-3}$] \\ \hline
$\rm oH_2D^+$ &  \,\,\,\,\,\,\,\,\,\,\,\,-\,\,\,\,\,\,\,\,\,\,\,\hspace{0.2mm} ($1_{11}$) & $\rm pH_2D^+$ & \,\,\,\,\,\,\,\,\,\,\,\,-\,\,\,\,\,\,\,\,\,\,\,\hspace{0.2mm} ($0_{00}$) & $\rm oD_2H^+$ & \,\,\,\,\,\,\,\,\,\,\,\,-\,\,\,\,\,\,\,\,\,\,\,\hspace{0.2mm} ($0_{00}$) & $\rm pD_2H^+$ & \,\,\,\,\,\,\,\,\,\,\,\,-\,\,\,\,\,\,\,\,\,\,\,\hspace{0.2mm} ($1_{01}$)\\
  &  $4.14\times10^5$ ($1_{10}$) & & $2.94\times10^6$ ($1_{01}$) & & $8.80\times10^6$ ($1_{11}$) & & $2.42\times10^6$ ($1_{10}$)\\
  &   $2.60\times10^7$ ($2_{12}$) & & $9.13\times10^7$ ($2_{02}$) & & $1.47\times10^7$ ($2_{02}$) & & $3.55\times10^7$ ($2_{12}$) \\
  &   $8.15\times10^7$ ($2_{11}$) & & $6.02\times10^8$ ($3_{03}$) & & $1.47\times10^7$ ($2_{11}$) & & $1.77\times10^8$ ($2_{21}$) \\
  &   $3.35\times10^8$ ($3_{13}$) & & $4.94\times10^7$ ($4_{04}$) & & $1.77\times10^8$ ($2_{20}$) & & $1.13\times10^7$ ($3_{03}$) \\
  &  $ 6.02\times10^8$ ($3_{12}$) & & $4.94\times10^7$ ($2_{21}$) & & $1.44\times10^8$ ($3_{13}$) & & $3.00\times10^9$ ($3_{12}$)\\

\hline
\end{tabular}
\label{tab1}
\end{table*}

\subsection{Species-to-species rate coefficients}\label{ss:spectospeccoeff}

Given that $\rm H_2D^+$ emission is observed to be ubiquitous (low-mass starless, pre-stellar and protostellar cores, \citealt{Caselli08}; high-mass star forming regions, \citealt{Pillai12}), it does not appear plausible to assume that the reactants lie only in their respective ground states in chemical reactions. However, up to now the effect of the higher-lying states on chemical reactions (for the $\rm H_3^+ + H_2$ system) has not been studied in the context of chemical modeling. It is this point that we want to study in the present paper.

Rate coefficients assuming that higher rotational levels can be populated -- the so-called species-to-species rate coefficients -- can be constructed from the state-to-state rate coefficients calculated by \citet{Hugo09}. Below, we show how this is accomplished in practice.

When the quantum states of the reactants and products are resolved, a chemical reaction can be written as
$$
{\rm A}_i + {\rm B}_j \arrow^{k_{ijmn}} {\rm C}_m + {\rm D}_n \; ,
$$ 
where the states are labeled with $i$, $j$, $m$, and $n$. The states considered in \citet{Hugo09} are the rotational levels of the $\hthree$ isotopologs and $\htwo$ in their ground vibrational states. The ground state-to-species coefficients pertain to the reaction 
$$
{\rm A_0} + {\rm B_0} \arrow^{k_{00}} {\rm C} + {\rm D} \; ,
$$ 
where the species A and B are in their ground states and C and D can enter into any state upon formation. The coefficient $k_{00}$ is obtained through summation over the possible product states:
$$
k_{00} = \sum_{m,n} k_{00mn} \; .
$$
The species-to-species rate coefficient, $\bar{k}$, is defined in terms of the total formation rates of C and D in reactions between A and B:
$$
{\bar k} [{\rm A}][{\rm B}] = \sum_{ij} \sum_{mn} k_{ijmn} 
[{\rm A}_i][{\rm B}_j] = \sum_{ij} k_{ij} \,
[{\rm A}_i][{\rm B}_j]\; ,
$$
where $[{\rm A}_i]$ is the number density (in units of cm$^{-3}$) of species A in state $i$, etc., and $k_{ij}$ is the sum over all the product states $m$ and $n$. The populations of the energy levels of A and B depend on the gas density and temperature, the cross-sections for collisional transitions, and the Einstein coefficients of the radiative transitions between the energy levels. For example, if we have a reason, based on these parameters, to believe that only the two lowest energy levels of A and only the ground state of B are populated, the species-to-species rate coefficient can be calculated from
$$
{\bar k} = \frac{k_{00}[{\rm A}_0][{\rm B}_0] + k_{10}[{\rm A}_1][{\rm B}_0]}{[{\rm A}][{\rm B}]} \; .
$$
At very high densities, or when the Einstein coefficients are very small, collisional excitation overpowers radiative transitions, and the level populations follow the Boltzmann distribution. In this case the species-to-species rate coefficient can be written as
$$
{\bar k} = \frac{\sum_{ij} k_{ij} \, g^{\rm A}_i e^{-E^{\rm A}_i/T} \, 
g^{\rm B}_j\,e^{-E^{\rm A}_j/T}}{Q^{\rm A}(T)\,Q^{\rm B}(T)} \; ,
$$
where $T$ is the kinetic temperature, $E^{\rm A}_i$ and $E^{\rm B}_j$ are the state energies (in K), $g^{\rm A}_i$ and $g^{\rm B}_j$ are the statistical weights of the levels (a product of the spin and rotational statistical weights), and $Q^{\rm A}(T)$ and $Q^{\rm B}(T)$ are the partition functions, 
$$
Q^{\rm A}(T) = \sum_i g^{\rm A}_i e^{-E^{\rm A}_i/T} \; , \;
Q^{\rm B}(T) = \sum_j g^{\rm B}_j e^{-E^{\rm B}_j/T} \; .
$$
The energy levels of the various rotational states, and the nuclear spin and rotational weights, can be read off Table~II in \citet{Hugo09}.

\begin{figure*}
\centering
\includegraphics[width=2.0\columnwidth]{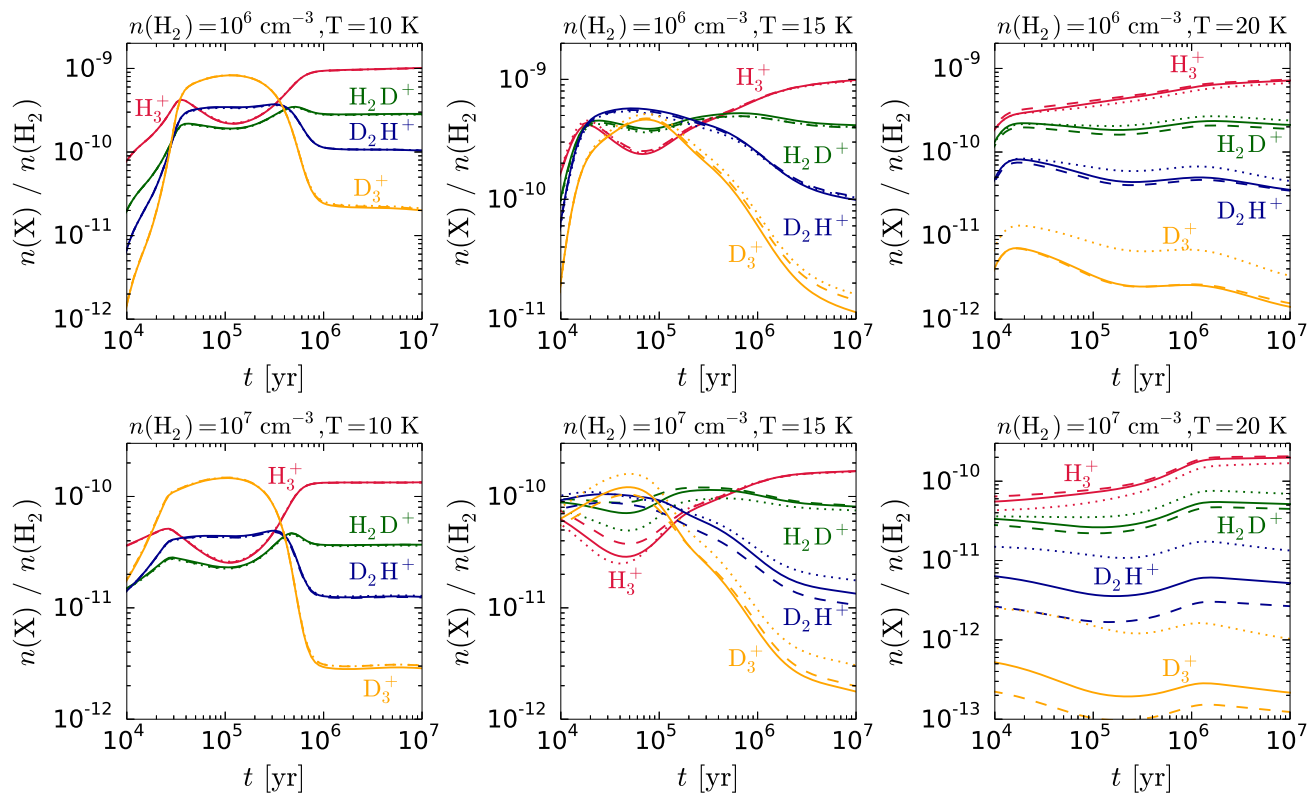}
\caption{Total abundances (sums over spin states) of the various $\rm H_3^+$ isotopologs as functions of time. The medium density is $n({\rm H_2}) = 10^6\,\rm cm^{-3}$ (upper row) or $n({\rm H_2}) = 10^7\,\rm cm^{-3}$ (lower row). From left to right, the panels show calculations assuming $T_{\rm gas} = T_{\rm dust} = 10\,$, 15, or 20\,K. Species-to-species rate coefficients are adopted in two of the models (method~1, dashed lines; method~2, solid lines). The dotted lines show the results of calculations using the ground state-to-species rate coefficients.
}
\label{fig:abundances}
\end{figure*}

The use of the species-to-species rate coefficients is only sensible if the medium density is high enough so that rotational states above the ground state can be assumed to be (significantly) populated, i.e., if the medium density is comparable to or above the critical density of a given rotational transition. Setting the appropriate values of the critical densities is not straightforward because they depend on the temperature. Furthermore, the relative strengths of collisional and radiative excitation change gradually as the density increases. In this work, we define the critical density as a value of medium density for which a rotationally excited level population is 0.8 times the value expected from the Boltzmann distribution. This choice is arbitrary, and its effect is discussed in Sect.\,\ref{s:discussion}. The level populations were determined using the radiative transfer program of \citet{Juvela97}. In this scheme the critical densities depend also on the abundances of the various species.

The calculation of the critical densities is discussed in greater detail in Appendix~\ref{a:coeffderiv}, where we present the critical densities of $\rm H_2D^+$ and $\rm D_2H^+$ as functions of temperature and abundance. Table~\ref{tab1} shows the critical densities of the various rotational energy levels of $\rm H_2D^+$ and $\rm D_2H^+$ adopted in this paper. These values have been collected from the tables given in Appendix~\ref{a:coeffderiv} and are shown here for convenience. Evidently, for a medium density of $n({\rm H_2}) = 10^6 \, \rm cm^{-3}$, for example, we can expect the lowest two rotational levels of $\rm oH_2D^+$ to be populated at low temperature\footnote{This is reinforced by the fact that the separation between the two lowest levels is only $\sim$20\,K \citep{Hugo09}.}, while $\rm pH_2D^+$, $\rm pD_2H^+$, and $\rm oD_2H^+$ should all lie mainly in their rotational ground states.

The final value of the species-to-species rate coefficient depends on which energy levels are taken into account. In this paper we consider two different approaches to choosing the included levels. We discuss these approaches next.

\begin{figure*}
\centering
\includegraphics[width=2.0\columnwidth]{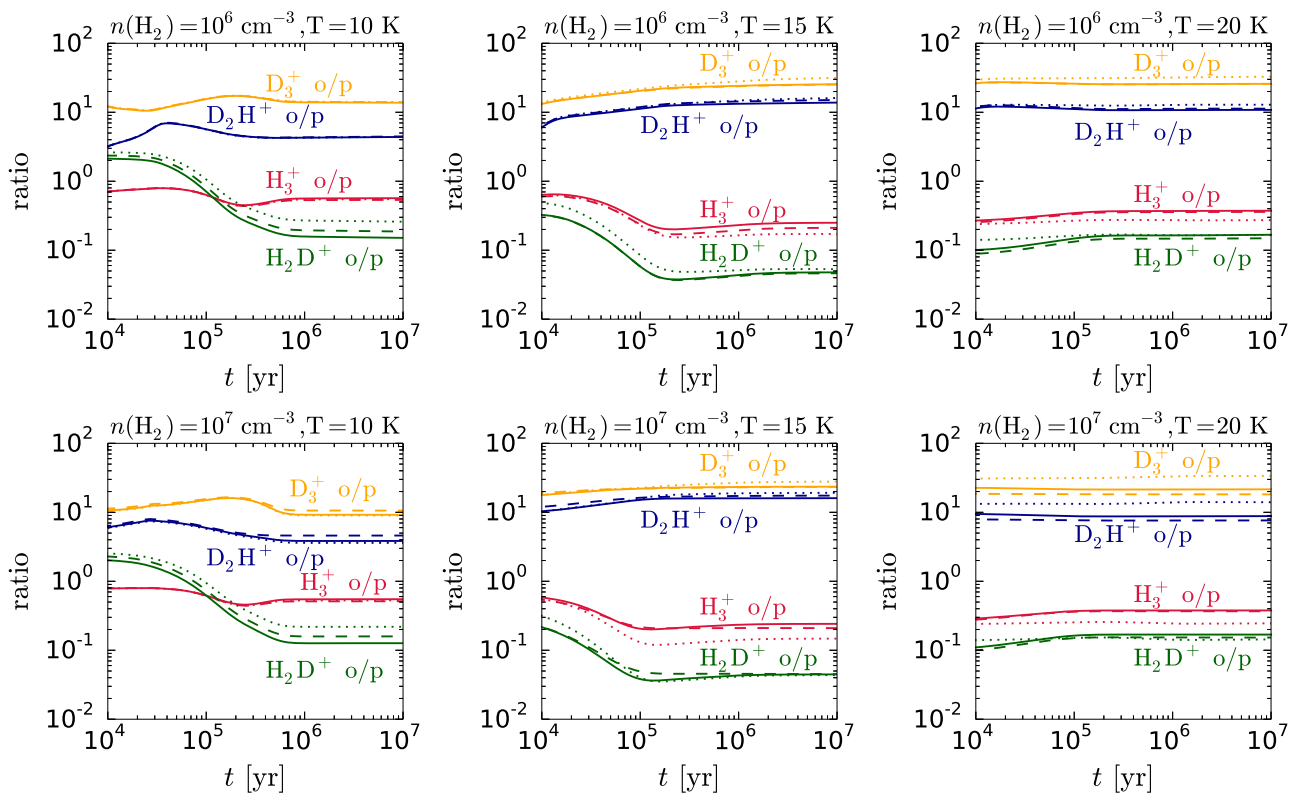}
\caption{Spin-state abundance ratios of the various $\rm H_3^+$ isotopologs as functions of time. The medium density is $n({\rm H_2}) = 10^6\,\rm cm^{-3}$ (upper row) or $n({\rm H_2}) = 10^7\,\rm cm^{-3}$ (lower row). From left to right, the panels show calculations assuming $T_{\rm gas} = T_{\rm dust} = 10\,$, 15, or 20\,K. Species-to-species rate coefficients are adopted in two of the models (method~1, dashed lines; method~2, solid lines). The dotted lines show the results of calculations using the ground state-to-species rate coefficients.
}
\label{fig:opratios}
\end{figure*}

\subsubsection{Local thermal equilibrium}

We consider first a scheme where all of the excited rotational states are accessible as long as the medium density is higher than the critical density of the first excited rotational state of (o,p)$\rm H_2D^+$ or (o,p)$\rm D_2H^+$. This situation corresponds to local thermodynamic equilibrium (LTE). One great advantage of this approach is that it allows the construction of a reaction set that can be easily read into a chemical model. We calculated the species-to-species rate coefficients for all reactions included in the $\rm H_3^+ + H_2$ reacting system and fitted the results with a modified Arrhenius rate law in the temperature range 5-50\,K. The resulting reaction set is given in Table~\ref{tabb2}. We stress that the rate coefficients given in this table are only applicable for (o,p)$\rm H_2D^+$ or (o,p)$\rm D_2H^+$ depending on the medium density as explained above. Because $\rm H_3^+$ and $\rm D_3^+$ are homonuclear molecules and do not have a permanent dipole moment, we assume that the species-to-species rate coefficients can be used at all medium densities for these species. In what follows, we refer to this LTE-based approach as ``method~1''.

\subsubsection{Restricted states}\label{sss:method2}

A more careful treatment of the (o,p)$\rm H_2D^+$ or (o,p)$\rm D_2H^+$ species-to-species rate coefficients involves selecting only those states that have a critical density below the medium density. Further restrictions apply: for example if the medium density is $n({\rm H_2}) = 5 \times 10^7 \, \rm cm^{-3}$, we include only the $1_{01}$, $1_{10}$, and $2_{12}$ rotational levels of $\rm pD_2H^+$ and not the $3_{03}$ level, even though it is allowed by the medium density, because the $2_{21}$ level is not accessible owing to our assumptions (see Table~\ref{tab1}). However, for $\rm H_3^+$ and $\rm D_3^+$ we again assume that all levels can be populated regardless of the medium density.

Because the values of the rate coefficients are now strictly tied to the density, the rate coefficients need to be calculated on a case-by-case basis and the construction of a ready-made reaction set is not practical. Instead, the calculation of the rate coefficients is performed internally in our chemical code. We call this restricted-state approach ``method~2''.

\section{Results}\label{s:results}

\subsection{Single-point models}

Figure~\ref{fig:abundances} shows the abundances (sums over spin states) of the $\rm H_3^+$ isotopologs as calculated with single-point chemical models assuming different values of medium density and temperature ($T_{\rm gas} = T_{\rm dust}$). Figure~\ref{fig:opratios} shows the spin-state abundance ratios in the same models. One feature of the models is immediately evident: the difference between methods~1~and~2 is small, i.e., one can employ the species-to-species rate coefficients given in Table~\ref{tabb2} with good confidence when modeling cold and dense environments. We checked that at $T = 50$\,K the difference between methods~1~and~2 remains smaller than a factor of two. However, at such a high temperature the abundances of the $\rm H_3^+$ isotopologs are so low that the chosen method is of no practical significance.

The total abundances of the $\rm H_3^+$ isotopologs are unaffected by the changes in the $\rm H_3^+ + H_2$ rate coefficients at $T = 10$\,K. The difference between the ground state-to-species model and the species-to-species model increases slightly with temperature, but more strongly with density, which is expected since the effect of the higher-lying rotational states becomes larger as consecutively higher states are populated. The general tendency in the species-to-species model is that the $\rm H_3^+$ deuteration degree decreases with respect to the ground state-to-species model toward higher temperatures and densities. This is due to the activation of several backward reaction channels that are suppressed when using the ground state-to-species coefficients. The simplistic approach of method~1 slightly underestimates the deuteration degree at $n({\rm H_2}) = 10^7\,\rm cm^{-3}$ and $T = 20$\,K. We discuss the rate coefficients of key reactions in Sect.\,\ref{s:discussion}.

The spin-state abundance ratios (Fig.\,\ref{fig:opratios}) are naturally also affected by the changes in the rate coefficients. At $n({\rm H_2}) = 10^6\,\rm cm^{-3}$ and $T = 10$\,K, only the ortho/para ratio of $\rm H_2D^+$ is modified, which is expected based on the critical densities (Table~\ref{tab1}) as o$\rm H_2D^+$ is the only species with accessible excited rotational states at $n({\rm H_2}) = 10^6\,\rm cm^{-3}$. Notably, the ortho/para ratios of $\rm H_3^+$ and $\rm D_3^+$ are almost unchanged at these conditions even though the species-to-species rate coefficients are assumed to be applicable at all densities for these two species. This is because of the large energy differences between the first excited rotational states and the ground states for $\rm H_3^+$ and $\rm D_3^+$ \citep{Hugo09}, which hinders rotational excitation at low temperature. The overall difference between the ground state-to-species and species-to-species models increases with temperature and density as was the case with the total abundances (Fig.\,\ref{fig:abundances}). However, the difference between the two models for any given spin-state abundance ratio is typically less than a factor of two.

\subsection{Source models: IRAS 16293 and L1544}

\begin{figure*}
\centering
\includegraphics[width=2.0\columnwidth]{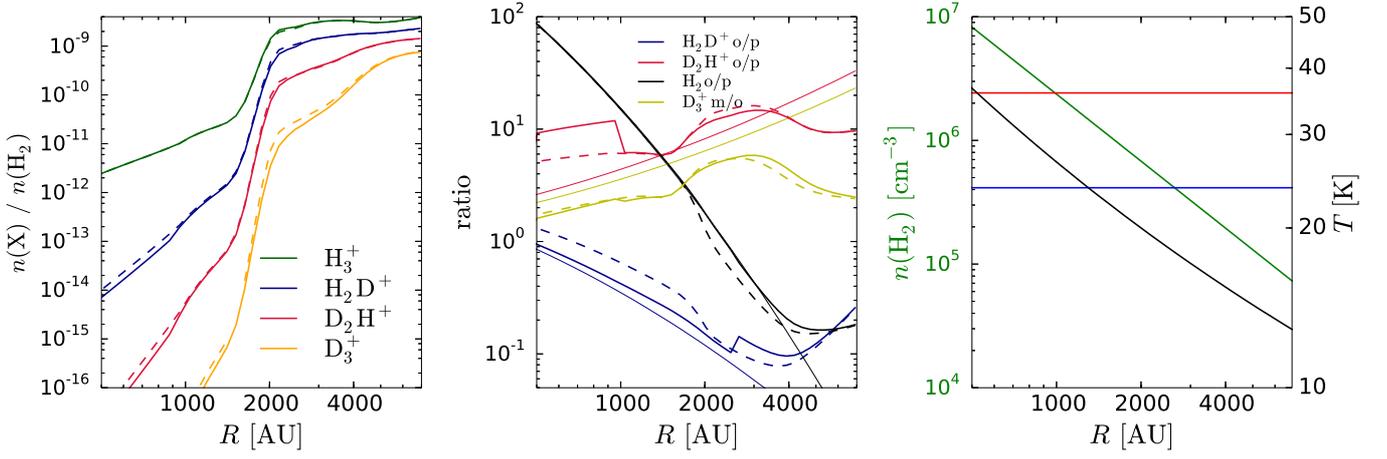}
\caption{{\sl Left:} Radial distributions of the fractional abundances of selected species at $t=5\times10^5$ yr in a protostellar core resembling IRAS 16293 according to the model of \citet{Crimier10}. Solid lines correspond to the species-to-species model (method~1), while dashed lines correspond to the ground state-to-species model. {\sl Middle:} Radial distributions of the o/p ratios of selected species, and the meta/ortho ratio of $\rm D_3^+$. The thin solid lines show the thermal spin-state ratios of the plotted species. {\sl Right:} Density and temperature distributions of the IRAS 16293 core model. The blue and red horizontal lines mark the critical densities of the first excited rotational transitions of $\rm oH_2D^+$ and $\rm pD_2H^+$, respectively.}
\label{fig:opratios_16293}    
\end{figure*}

\begin{figure*}
\centering
\includegraphics[width=2.0\columnwidth]{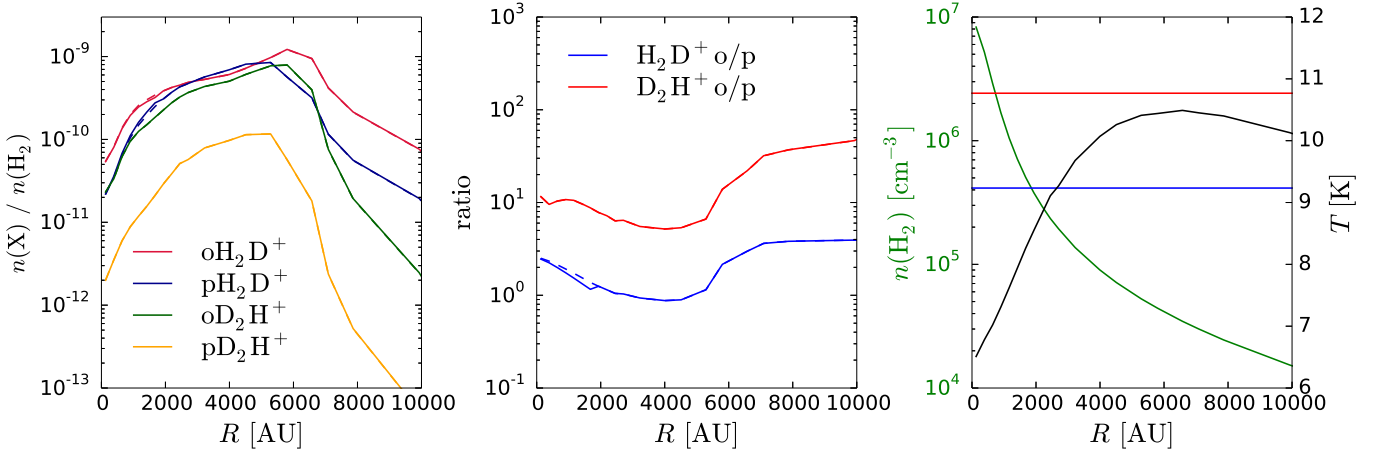}
\caption{{\sl Left:} Radial distributions of the fractional abundances of the spin states of $\rm H_2D^+$ and $\rm D_2H^+$ at $t=1\times10^6$ yr in the innermost 10000\,AU of the L1544 model. Solid lines correspond to the species-to-species model (method~1), while dashed lines correspond to the ground state-to-species model. {\sl Middle:} Radial distributions of the o/p ratios of $\rm H_2D^+$ and $\rm D_2H^+$. Linestyles are the same as in the left panel. {\sl Right:} Density and temperature distributions of the L1544 core model. The blue and red horizontal lines mark the critical densities of the first excited rotational transitions of $\rm oH_2D^+$ and $\rm pD_2H^+$, respectively.}
\label{fig:L1544}
\end{figure*}

It is not obvious based on the single-point models how and if the abundances and spin-state ratios of the $\rm H_3^+$ isotopologs change in the context of source models with radially-varying density and temperature profiles, when one switches from ground state-to-species rate coefficients to species-to-species rate coefficients. Potential problems arising from the use of ground state-to-species rate coefficients for the $\rm H_3^+ + H_2$ system in the interpretation of observational data is discussed by \citet{Harju17b}. Following the discussion in that paper, we carried out chemical calculations using a physical model for the protostellar system IRAS 16293-2422 A/B (\citealt{Crimier10}; see also \citealt{Brunken14}; \citealt{Harju17b}). The chemical modeling setup is essentially the same as described in \citet{Brunken14}, where the physical model is also described in detail. The source model, which assumes spherical symmetry, is separated into concentric shells and the chemical evolution in each shell is tracked using our pseudo-time-dependent chemical code (Sect.\,\ref{ss:chemmodel}). The model outputs the abundances of the various species as functions of radius and time, allowing us to reconstruct the time-evolution of the $\rm H_3^+$ isotopologs (among others) in the model core. We carried out chemical calculations using the ground state-to-species and species-to-species rate coefficients for the $\rm H_3^+ + H_2$ system and compared the difference in predicted radial abundance profiles for the $\rm H_3^+$ isotopologs and specifically the o/p ratios of $\rm H_2D^+$ and $\rm D_2H^+$.

\begin{figure*}
\centering
\includegraphics[width=2.0\columnwidth]{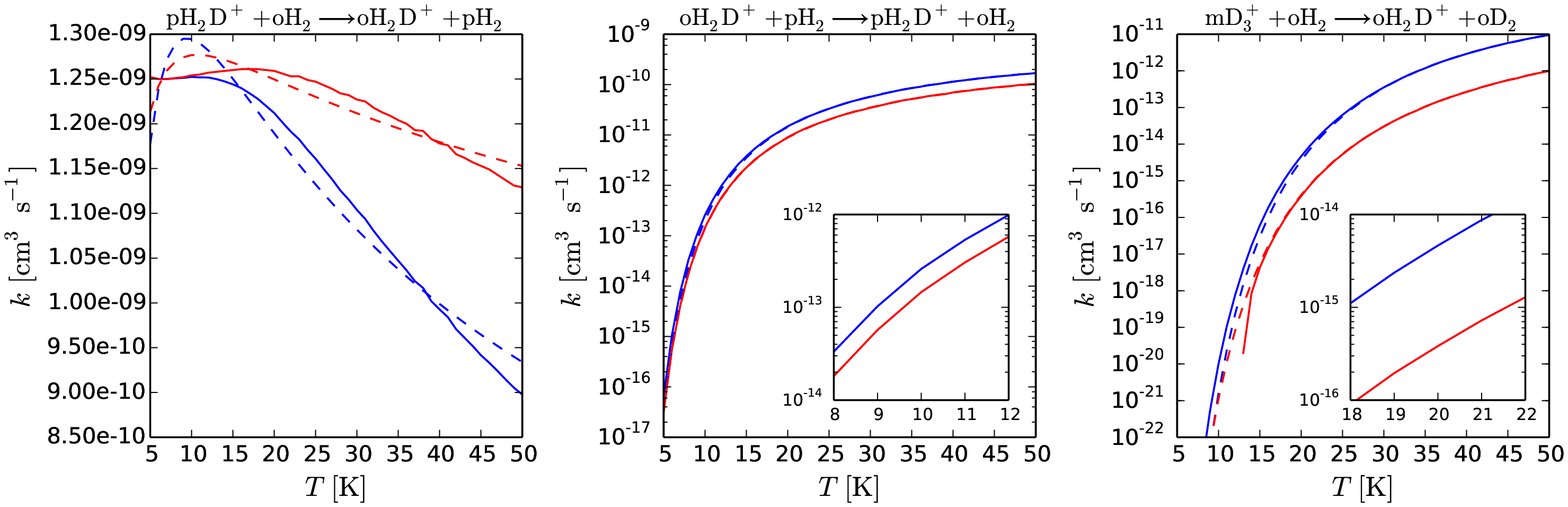}
\caption{Rate coefficients of the $\rm pH_2D^+ + oH_2 \longrightarrow oH_2D^+ + pH_2$ ({\sl left}), $\rm oH_2D^+ + pH_2 \longrightarrow pH_2D^+ + oH_2$ ({\sl middle}), and $\rm mD_3^+ + oH_2 \longrightarrow oH_2D^+ + oD_2$ ({\sl right}) reactions as functions of temperature. Red lines represent ground state-to-species rate coefficients, while blue lines represent species-to-species rate coefficients. Solid lines represent the raw data from \citet{Hugo09}; dashed lines represent our fits using the modified Arrhenius rate law (see text). Note the different y-axis scaling in the panels.
}
\label{fig:ratecoeffs}
\end{figure*}

Figure~\ref{fig:opratios_16293} shows the total abundances (summed over spin states) of the $\rm H_3^+$ isotopologs and the o/p ratios of $\rm H_2D^+$, $\rm D_2H^+$, and $\rm H_2$, and the meta/ortho ratio of $\rm D_3^+$, at $t = 5.0 \times 10^5\,\rm yr$ in the IRAS 16293 model. The figure concentrates on the radius range where most of the $\rm H_2D^+$ and $\rm D_2H^+$ emission/absorption originates (\citealt{Brunken14}; \citealt{Harju17b}). The total abundances are not significantly affected by the choice of the $\rm H_3^+ + H_2$ rate coefficients, but the spin-state abundance ratios show differences depending on the rate coefficients used. In particular, we recover exactly the type of behavior expected on the basis of the discussion in \citet{Harju17b}: the $\rm H_2D^+$ o/p ratio decreases by a factor of $\sim$1.5 and the $\rm D_2H^+$ o/p ratio increases by a similar factor when one switches from the ground state-to-species to species-to-species rate coefficients. The sharp features in the o/p ratios evident at $R \sim 2500\,\rm AU$ for $\rm H_2D^+$ and at $R \sim 1000\,\rm AU$ for $\rm D_2H^+$ are a result of the activation of the species-to-species rate coefficients at the densities corresponding to these radii (Table~\ref{tab1}). We note that for $\rm H_2D^+$ for example the solid and dashed lines do not overlap below the critical density of the first excited rotational state, even though the ground state-to-species rate coefficients apply in this regime. This is because of the effect of $\rm H_3^+$ and $\rm D_3^+$, for which the species-to-species rate coefficients apply in all conditions.

To investigate the effect of the species-to-species rates in prestellar core conditions, we performed another test using the physical model for the prestellar core L1544 discussed in \citet{Keto10} and \citet{Keto14}. The chemical calculation procedure was identical to that used in the IRAS 16293 modeling. Figure~\ref{fig:L1544} shows the results of the calculations at $t = 10^6\,\rm yr$, zoomed in to the innermost 10000\,AU of the core model. The density is high enough only in the central 2000\,AU to activate the species-to-species rate coefficients for $\rm oH_2D^+$, while $\rm pD_2H^+$ is only affected in the central $\sim$500\,AU. The results from the two types of model are however more or less identical, which is caused by the low temperature in the central regions. This test confirms the tendencies shown in Fig.\,\ref{fig:abundances}, i.e., that the species-to-species rate coefficients become important at high density only if the temperature is higher than 10\,K.

\section{Discussion}\label{s:discussion}

The results presented in Sect.\,\ref{s:results} clearly show that the switch from ground state-to-species rate coefficients to species-to-species rate coefficients can have a marked impact on the chemistry of the $\rm H_3^+$ isotopologs. As discussed in \citet{Brunken14} and \citet{Harju17b}, the $\rm pH_2D^+ + oH_2 \leftrightarrow oH_2D^+ + pH_2$ reaction, endothermic in the backward direction, plays a major role in the evolution of the $\rm H_2D^+$ o/p ratio. Figure~\ref{fig:ratecoeffs} shows the rate coefficient of this reaction in the forward and backward directions. The ground state-to-species model slightly overestimates the destruction of $\rm pH_2D^+$ at $T > 10$\,K, although the difference between the two cases is only $\sim$20\%. The backward rate coefficient is however clearly higher in the species-to-species model (calculated with method~1) throughout the temperature range considered; the difference is about a factor of 1.8 at $T = 10$\,K, and $\sim$1.5-2 throughout the temperature range considered. This difference in the backward rate coefficient -- although only of a factor of two -- can translate to a decrease in the $\rm H_2D^+$ o/p ratio as evident in the figures presented in Sect.\,\ref{s:results}, and can also be observable \citep{Harju17b}.

For many reactions pertaining to the $\rm H_3^+ + H_2$ system the ground state-to-species and species-to-species rate coefficients are very similar, showing maximum differences on the order of 10\%. The general tendency is that large differences are usually seen in the rates of endothermic reactions, which are underestimated by the ground state-to-species model, as displayed in Fig.\,\ref{fig:ratecoeffs}. The $\rm mD_3^+ + oH_2 \longrightarrow oH_2D^+ + oD_2$ reaction (shown in the right panel in Fig.\,\ref{fig:ratecoeffs}) was included to demonstrate that sometimes the difference between the rate coefficients may be larger than one order of magnitude (for this reaction about a factor of 12 at 20\,K). It is clear from the above that one should employ the species-to-species rate coefficients in a chemical model to obtain a better estimate of the abundances of the the spin states of the $\rm H_3^+$ isotopologs.

The modeling estimates for the abundances of the $\rm H_3^+$ isotopologs are tied to the critical densities, as this affects when the species-to-species rate coefficients are assumed to apply. The critical densities depend relatively strongly on the abundances of the species at high density where optical depth effects become important. However, it is indeed the abundances that we wish to obtain from the chemical modeling. To derive truly consistent values for the various critical densities, one should iterate the chemical calculations in order to obtain successively better estimates for the critical densities and, in turn, the abundances. In this paper we settled on a simple scheme where the critical densities (Table~\ref{tab1}) were chosen on the basis of steady-state abundances obtained from our rate coefficient fitting test (Fig.\,\ref{fig:gstscomparison}).

The choice of the cutoff for when the species-to-species rates are assumed apply (see Appendix~\ref{a:coeffderiv} for more details) is arbitrary, and it is understood that the use of species-to-species rate coefficients adds to the parameter space in the modeling, and so is an additional source of uncertainty. Altering the critical densities would affect the radii where the discontinuities in Figs.\,\ref{fig:opratios_16293}~and~\ref{fig:L1544} appear. Here we considered that a rotationally excited level is taken into account in the calculation of the rate coefficient if its population is $\gtrsim 0.8$ times that of the thermal population. The goodness of this choice should be tested through a comparison against a state-to-state chemistry model, which is, however, beyond the scope of the present paper. The species that is the most affected by the choice of the threshold value is $\rm oH_2D^+$. Using a threshold ratio of 0.9 would shift the discontinuity from 2500\,AU to 1500\,AU, whereas with a threshold of 0.7 the discontinuity would appear at a distance of 3500\,AU from the center. The threshold 0.9 gives practically the same abundance profile as the ground state-to-species model does. The mentioned changes alter the line-of-sight average abundance of $\rm oH_2D^+$ less than 10 percent. Combined with the steep density and temperature gradients of this model, the small changes in the $\rm oH_2D^+$ abundance profile have, however, a noticeable effect on the observed line intensity. We note that the uncertainty associated with the threshold is not likely to be greater than, for example, the uncertainty in the temperature in the modeled objects (typically $\sim$1-2 K for the cold gas). To summarize the experiences from the present study, an accurate model of the chemistry of the $\rm H_3^+$ isotopologs should include the effect of excited rotational states, but the outcome also depends on how well the density and temperature distributions of the object are known.

\section{Conclusions}\label{s:conclusions}

We studied the chemistry of the spin states of the $\rm H_3^+$ isotopologs using a model where rotational excitation of the reactants is taken into account (the species-to-species model), as opposed to previous studies of the subject which have thus far assumed that only the rotational ground state is populated (the ground state-to-species model). We considered two different methods to constructing the species-to-species rate coefficients: the first method assumes that all rotational states can be populated once the medium density is higher than the critical density of the first excited rotational state, while in the second method only those rotational states whose critical densities lie below the medium density are included.

We found that the two methods of constructing the species-to-species rate coefficients produce similar results in the physical conditions considered here, which means that the species-to-species rate coefficients can be conveniently read from a precalculated table with good confidence, and are thus easily implementable in a chemical model. On the other hand, the abundances and spin-state ratios predicted by the ground state-to-species and species-to-species models are different from each other at high density, depending also on the temperature. Notable differences between the models are seen for $T > 10$\,K where the ground state-to-species model overestimates the abundances of the $\rm H_3^+$ isotopologs, by a factor of $\sim$2 for $\rm H_2D^+$ and even up to an order of magnitude for $\rm D_3^+$. The species-to-species model is the more realistic one of the two. We note however that the species-to-species rate coefficients introduce an additional uncertainty into the modeling through the critical densities of the rotational transitions of $\rm H_2D^+$ and $\rm D_2H^+$, which determine when the species-to-species model is applicable.

The new model is a step toward full state-to-state modeling of the $\rm H_3^+$ isotopolog chemistry. A state-to-state model would not suffer from the problem of setting the appropriate values of critical densities like the present model does, but the implementation of state-to-state rates into a complete gas-grain chemical model including elements heavier than hydrogen is a monumental task. Such an effort, although certainly called for, is left for future work. In the meantime, the species-to-species rate coefficients represent the next best thing, and should be implemented in models of the chemistry of the $\rm H_3^+ + H_2$ reacting system.

\begin{acknowledgements}
P.C. acknowledges financial support of the European Research Council (ERC; project PALs 320620).
\end{acknowledgements}

\bibliographystyle{aa}
\bibliography{h3+.bib}

\onecolumn

\appendix

\newpage

\section{Determining the critical densities}\label{a:coeffderiv}

\begin{figure*}[bt]
\unitlength=1mm
\begin{picture}(160,130)(0,0)

\put(0,65){
\begin{picture}(0,0) 
\includegraphics[width=9cm, angle=0]{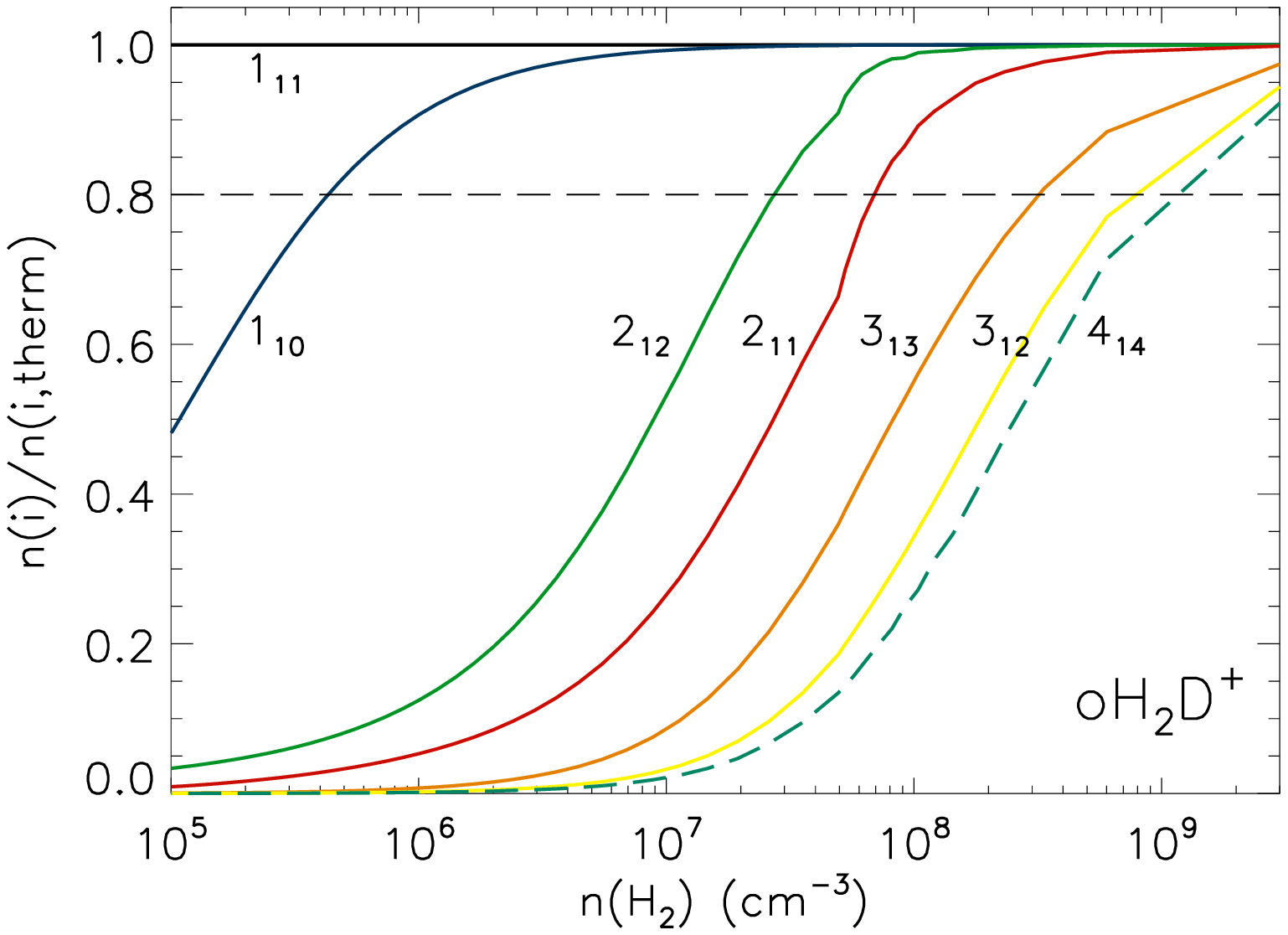}
\end{picture}}

\put(90,65){
\begin{picture}(0,0) 
\includegraphics[width=9cm, angle=0]{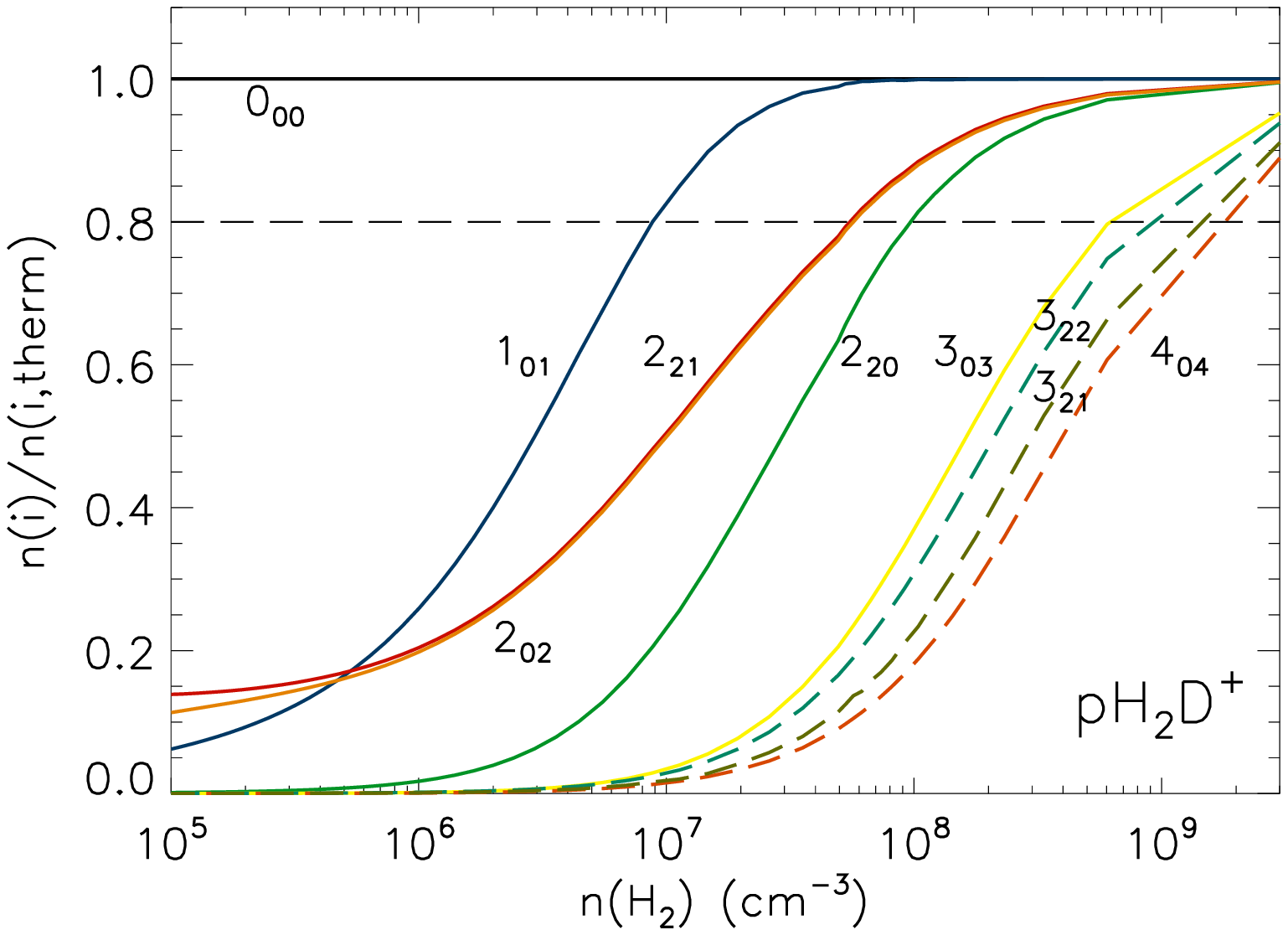}
\end{picture}}

\put(0,0){
\begin{picture}(0,0) 
\includegraphics[width=9cm,angle=0]{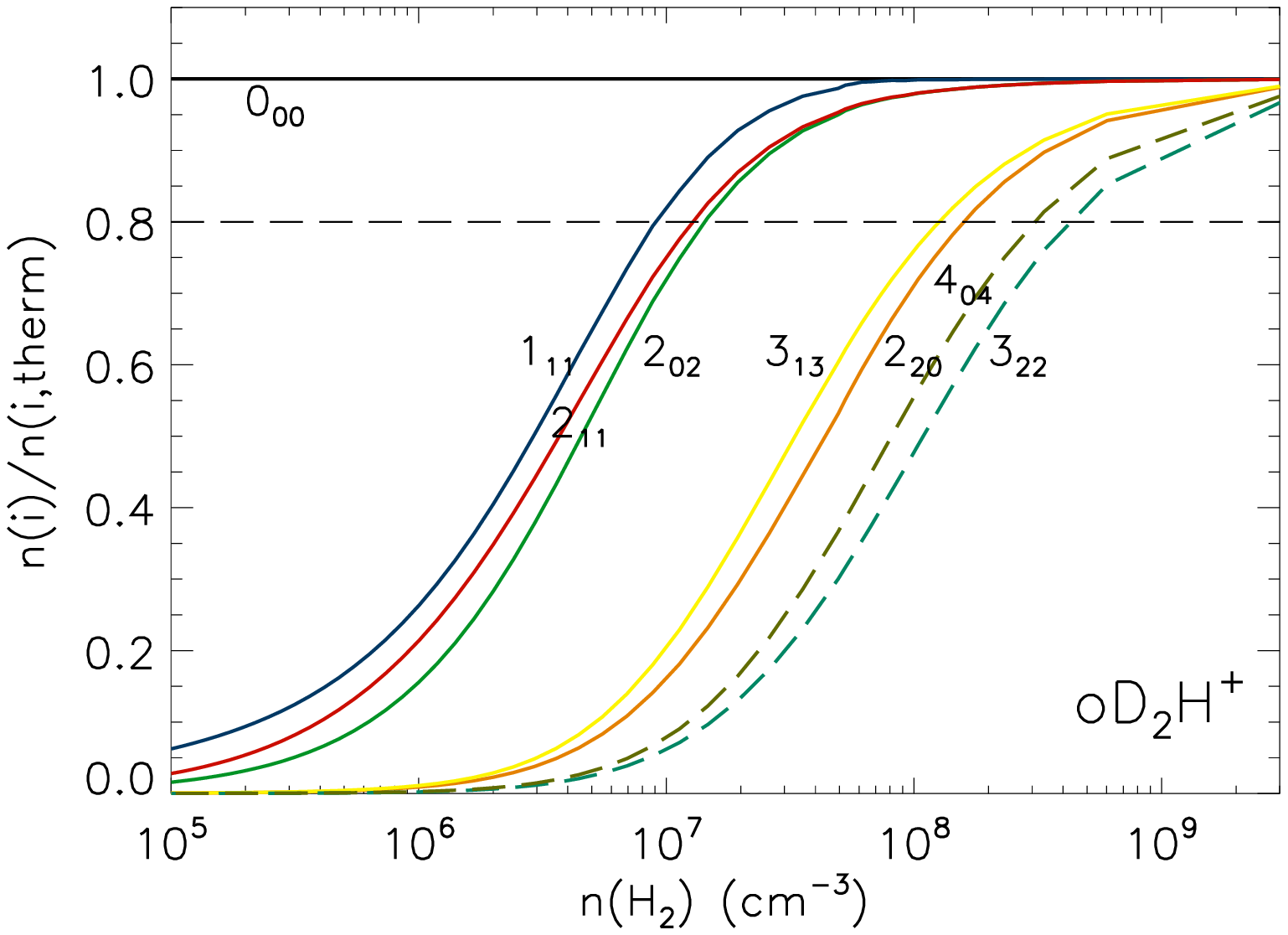}
\end{picture}}

\put(90,0){
\begin{picture}(0,0) 
\includegraphics[width=9cm,angle=0]{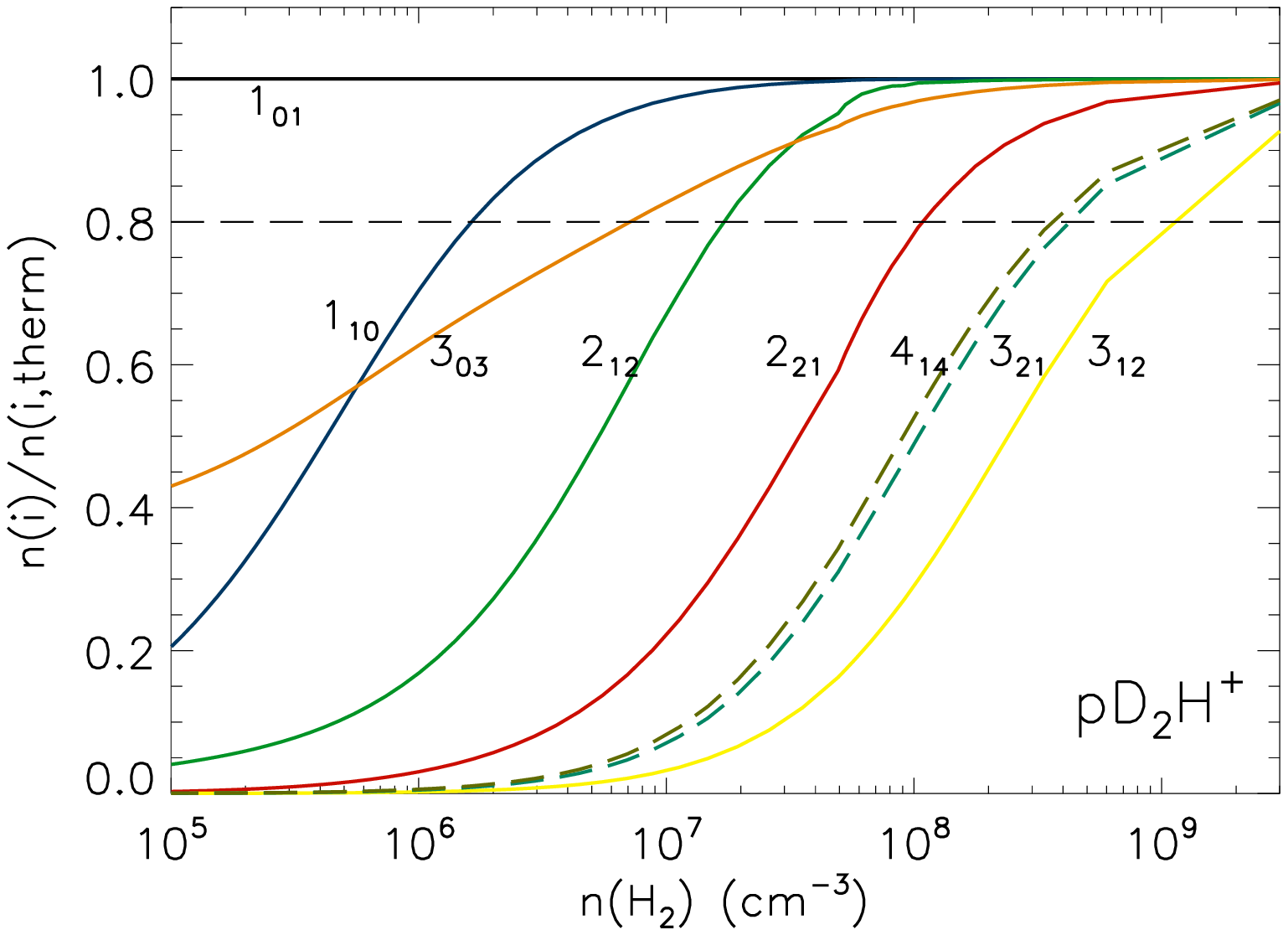}
\end{picture}}

\end{picture}
\caption{Excitation of the lowest rotational levels of o$\htwod$, p$\htwod$, o$\dtwoh$, and p$\dtwoh$ as functions of the gas density at $T = 15$\,K, assuming a fractional abundance $X=10^{-10}$ for each species. The curves show the ratios of the populations to the values expected from the Boltzmann distribution. The dashed horizontal lines indicates the limit where a level is considered to be significantly populated and taken into account in the calculation of the species-to-species rate coefficients.}
\label{figure:ncrits}
\end{figure*}

In the present study, the densities where the rotational levels of $\htwod$ and $\dtwoh$ become significantly populated were estimated
through radiative transfer calculations. The core model had a steep density gradient, increasing from $n(\htwo) = 10^5$ cm$^{-3}$ at the
outer boundary to $3\times10^9$ cm$^{-3}$ in the center. The core was assumed to be isothermal, and we varied the temperature in the range $5-25$~K. Different fractional abundances were also tested to examine optical thickness effects. The Monte Carlo program of \citet{Juvela97} was used in the simulations. The results for the temperature $T=15$\,K are shown in Fig.\,\ref{figure:ncrits}. The curves show the ratios of the populations to the values expected from the Boltzmann distribution. In the figure, the fractional abundances of all the four species, o$\htwod$, p$\htwod$, o$\dtwoh$, and p$\dtwoh$, are assumed to be $X=10^{-10}$. A level is thermalized when the ratio is unity. In ``method 2'' of the chemistry model described in \ref{sss:method2}, a level is taken into account, and its population relative to the ground level is calculated by the Boltzmann factor, at densities where the ratio to the thermal value exceeds 0.8. This limit is shown with dashed horizontal lines in Fig.\,\ref{figure:ncrits}. The critical densities of o$\htwod$, p$\htwod$, o$\dtwoh$, and p$\dtwoh$ calculated with the radiative transfer model for different temperatures and for different abundances of the species are collected in Tables~\ref{taba1}~to~\ref{taba3}, from which the values used in our model calculations, given in Table~\ref{tab1}, were compiled.

\begin{figure*}[bt]
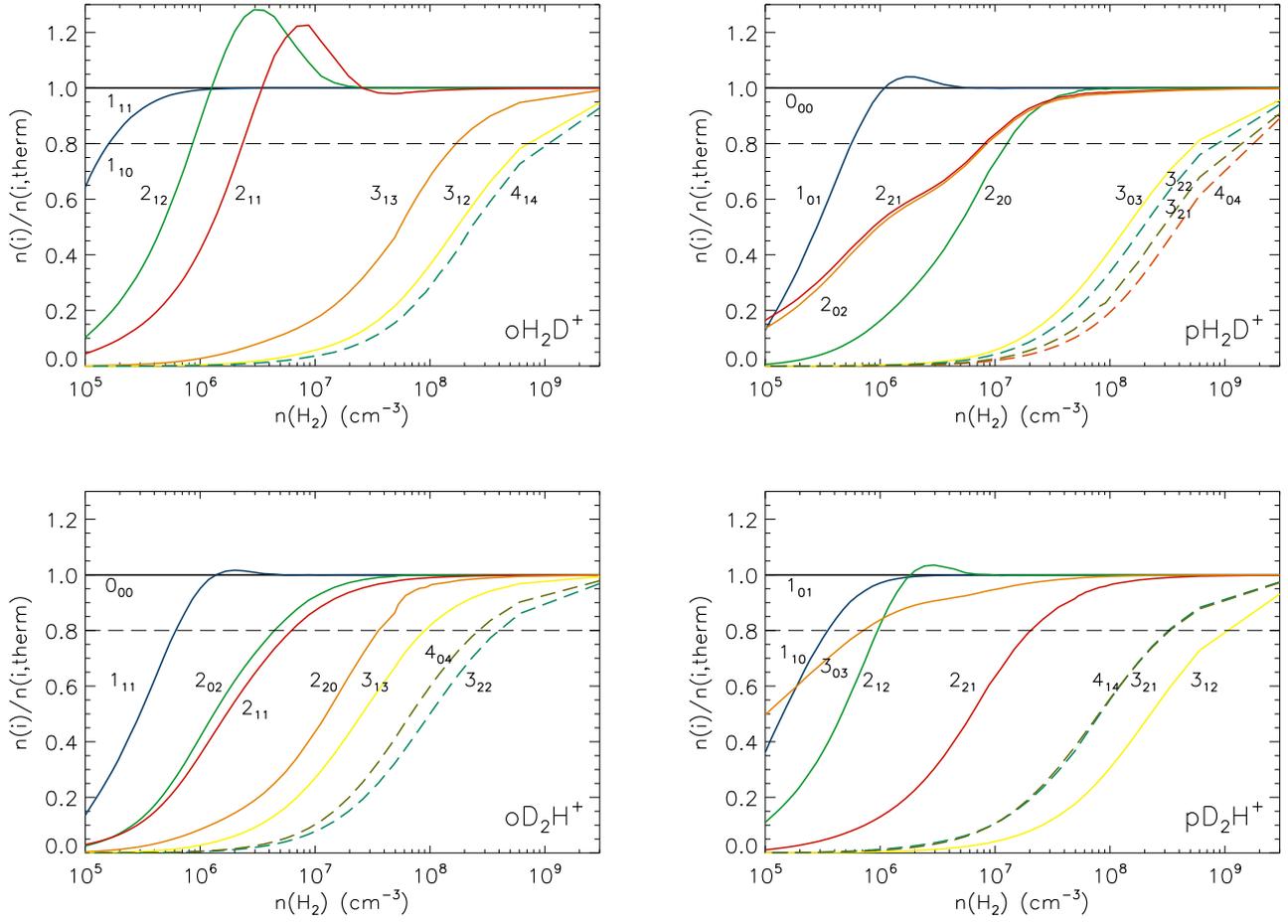

\unitlength=1mm

\caption{As Fig.\,\ref{figure:ncrits}, but assuming $X=10^{-8}$ for each species.}
\label{figure:ncrits_x-8}
\end{figure*}

For completeness we show in Fig.\,\ref{figure:ncrits_x-8} similar results as in Fig.\,\ref{figure:ncrits}, but assuming $X=10^{-8}$ for $\rm H_2D^+$ and $\rm D_2H^+$, which can be considered as a strong upper limit as the abundances are in reality unlikely to rise this high. The increase in the fractional abundance increases the optical thickness of the rotational transitions. The effect is that photons are absorbed and re-emitted several times before escaping the cloud, and the role of collisional transitions in determining the level populations becomes more important. This so called 'radiative trapping' influences the level populations in the same way as an increase in the gas density, and results in a greater thermalisation of the populations \citep{Walmsley87}. Comparison between Figs.\,\ref{figure:ncrits} and \ref{figure:ncrits_x-8} shows that assuming fractional abundances of $X=10^{-8}$ causes the thermalization of the lowest rotational levels to occur at clearly lower densities than in the case of $X=10^{-10}$. The levels $2_{12}$ and $2_{11}$ of $\rm oH_2D^+$ become mildly supra-thermally excited (the maximum excitation temperatures are $T_{\rm ex} \sim 15.4-15.5$ K) when $X=10^{-8}$ is assumed. This is caused by rapid radiative decay from the collisionally excited higher rotational levels. We note that for canonical abundances ($X=10^{-11}-10^{-9}$) supra-thermal populations are never found for rotationally excited levels of $\rm H_2D^+$ and $\rm D_2H^+$.

\newpage

\begin{table}
\caption{Critical densities ($n_{\rm c}$) of the rotational levels of $\rm H_2D^+$ and $\rm D_2H^+$, in order of increasing energy \citep{Hugo09} assuming $T = 10$\,K. The corresponding rotational level ($J_{K_aK_c}$) is shown in parentheses after each value of critical density. The three columns give the critical densities for different values of the fractional abundance $X$ of each species.}
\centering
\begin{tabular}{c c c c}
\hline \hline 
$\rm oH_2D^+$ & $X=10^{-11}$ & $X=10^{-10}$ & $X=10^{-9}$ \\ 
\hline 
& $5.48\times10^5~(1_{10})$ & $4.14\times10^5~(1_{10})$ & $2.80\times10^5~(1_{10})$ \\ 
& $5.69\times10^7~(2_{12})$ & $2.60\times10^7~(2_{12})$ & $8.80\times10^6~(2_{12})$ \\ 
& $2.32\times10^8~(2_{11})$ & $8.15\times10^7~(2_{11})$ & $3.55\times10^7~(2_{11})$ \\ 
& $6.02\times10^8~(3_{13})$ & $3.35\times10^8~(3_{13})$ & $3.35\times10^8~(3_{13})$ \\ 
& $3.00\times10^9~(3_{12})$ & $6.02\times10^8~(3_{12})$ & $6.02\times10^8~(3_{12})$ \\ 
\hline 
$\rm pH_2D^+$ & $X=10^{-11}$ & $X=10^{-10}$ & $X=10^{-9}$ \\ 
\hline 
& $1.94\times10^7~(1_{01})$ & $8.80\times10^6~(1_{01})$ & $2.94\times10^6~(1_{01})$ \\
\hline 
\end{tabular}
\label{taba1}
\end{table}

\begin{table}
\caption{Critical densities ($n_{\rm c}$) of the rotational levels of $\rm H_2D^+$ and $\rm D_2H^+$, in order of increasing energy \citep{Hugo09} assuming $T = 15$\,K. The corresponding rotational level ($J_{K_aK_c}$) is shown in parentheses after each value of critical density. The three columns give the critical densities for different values of the fractional abundance $X$ of each species.}
\centering
\begin{tabular}{c c c c}
\hline \hline 
$\rm oH_2D^+$ & $X=10^{-11}$ & $X=10^{-10}$ & $X=10^{-9}$ \\ 
\hline 
& $5.48\times10^5~(1_{10})$ & $4.75\times10^5~(1_{10})$ & $3.18\times10^5~(1_{10})$ \\ 
& $6.70\times10^7~(2_{12})$ & $3.55\times10^7~(2_{12})$ & $8.80\times10^6~(2_{12})$ \\ 
& $1.77\times10^8~(2_{11})$ & $7.35\times10^7~(2_{11})$ & $3.55\times10^7~(2_{11})$ \\ 
& $6.02\times10^8~(3_{13})$ & $3.35\times10^8~(3_{13})$ & $3.35\times10^8~(3_{13})$ \\ 
& $3.00\times10^9~(3_{12})$ & $3.00\times10^9~(3_{12})$ & $3.00\times10^9~(3_{12})$ \\ 
& $3.00\times10^9~(4_{14})$ & $3.00\times10^9~(4_{14})$ & $3.00\times10^9~(4_{14})$ \\ 
\hline 
$\rm pH_2D^+$ & $X=10^{-11}$ & $X=10^{-10}$ & $X=10^{-9}$ \\
\hline 
\end{tabular}
\label{taba2}
\end{table}

\begin{table}
\caption{Critical densities ($n_{\rm c}$) of the rotational levels of $\rm H_2D^+$ and $\rm D_2H^+$, in order of increasing energy \citep{Hugo09} assuming $T = 20$\,K. The corresponding rotational level ($J_{K_aK_c}$) is shown in parentheses after each value of critical density. The three columns give the critical densities for different values of the fractional abundance $X$ of each species.}
\centering
\begin{tabular}{c c c c}
\hline \hline 
$\rm oH_2D^+$ & $X=10^{-11}$ & $X=10^{-10}$ & $X=10^{-9}$ \\ 
\hline 
& $5.48\times10^5~(1_{10})$ & $4.75\times10^5~(1_{10})$ & $3.62\times10^5~(1_{10})$ \\ 
& $6.70\times10^7~(2_{12})$ & $3.55\times10^7~(2_{12})$ & $1.13\times10^7~(2_{12})$ \\ 
& $1.77\times10^8~(2_{11})$ & $6.70\times10^7~(2_{11})$ & $2.60\times10^7~(2_{11})$ \\ 
& $6.02\times10^8~(3_{13})$ & $3.35\times10^8~(3_{13})$ & $2.32\times10^8~(3_{13})$ \\ 
& $3.00\times10^9~(3_{12})$ & $3.00\times10^9~(3_{12})$ & $3.00\times10^9~(3_{12})$ \\ 
& $3.00\times10^9~(4_{14})$ & $3.00\times10^9~(4_{14})$ & $3.00\times10^9~(4_{14})$ \\ 
\hline 
$\rm pH_2D^+$ & $X=10^{-11}$ & $X=10^{-10}$ & $X=10^{-9}$ \\
\hline 
\end{tabular}
\label{taba3}
\end{table}

\newpage

\twocolumn

\newpage

\onecolumn

\section{Ground state-to-species and species-to-species rate coefficients}\label{a:tables}

\begin{longtable}{cccccccccc}
\caption{Ground state-to-species rate coefficients for the $\rm H_3^+ + H_2$ system calculated from the state-to-state data by \citet{Hugo09}. The fit parameters $\alpha$, $\beta$, and $\gamma$ correspond to the modified Arrhenius rate law $k = \alpha \, (T/300)^\beta \, \exp(-\gamma/T)$. The fit is valid in the temperature range 5-50\,K. Reactions forbidden by spin selection rules and identity reactions were omitted. The notation $a(b)$ denotes $a\times10^b$.}\\
\hline\hline
\centering

& & Chemical reaction & & & & $\alpha$ & $\beta$ & $\gamma$ \\ \hline
\endfirsthead
\caption{continued.}\\
\hline\hline
& & Chemical reaction & & & & $\alpha$ & $\beta$ & $\gamma$ \\ \hline
\endhead
$\rm pH_3^+$ & $\rm pH_2  $ & $\longrightarrow$  & $\rm pH_3^+$ & $\rm oH_2  $ & & 6.29(-10) & $-0.27$ & $174.90$ \\ 
$\rm pH_3^+$ & $\rm pH_2  $ & $\longrightarrow$  & $\rm oH_3^+$ & $\rm oH_2  $ & & 5.27(-10) & $-0.11$ & $203.05$ \\ 
$\rm pH_3^+$ & $\rm oH_2  $ & $\longrightarrow$  & $\rm pH_3^+$ & $\rm pH_2  $ & & 2.06(-10) & $-0.15$ & $0.73$ \\ 
$\rm pH_3^+$ & $\rm oH_2  $ & $\longrightarrow$  & $\rm oH_3^+$ & $\rm pH_2  $ & & 1.87(-10) & $-0.24$ & $1.36$ \\ 
$\rm pH_3^+$ & $\rm oH_2  $ & $\longrightarrow$  & $\rm oH_3^+$ & $\rm oH_2  $ & & 5.85(-10) & $-0.16$ & $35.24$ \\ 
$\rm pH_3^+$ & $\rm HD    $ & $\longrightarrow$  & $\rm oH_3^+$ & $\rm HD    $ & & 6.31(-11) & $0.06$ & $30.92$ \\ 
$\rm pH_3^+$ & $\rm HD    $ & $\longrightarrow$  & $\rm pH_2D^+$ & $\rm pH_2  $ & & 1.49(-10) & $-0.29$ & $1.65$ \\ 
$\rm pH_3^+$ & $\rm HD    $ & $\longrightarrow$  & $\rm pH_2D^+$ & $\rm oH_2  $ & & 2.72(-10) & $-0.24$ & $2.92$ \\
\end{longtable}

\newpage

\begin{longtable}{cccccccccc}
\caption{Species-to-species rate coefficients for the $\rm H_3^+ + H_2$ system calculated from the state-to-state data by \citet{Hugo09}. The fit parameters $\alpha$, $\beta$, and $\gamma$ correspond to the modified Arrhenius rate law $k = \alpha \, (T/300)^\beta \, \exp(-\gamma/T)$. The fit is valid in the temperature range 5-50\,K, and for a range of medium densities limited by the critical density (see the main text). Reactions forbidden by spin selection rules and identity reactions were omitted. The notation $a(b)$ denotes $a\times10^b$.}\\
\hline\hline
\centering
& & Chemical reaction & & & & $\alpha$ & $\beta$ & $\gamma$ \\ \hline
\endfirsthead
\caption{continued.}\\
\hline\hline
& & Chemical reaction & & & & $\alpha$ & $\beta$ & $\gamma$ \\ \hline
\endhead
$\rm pH_3^+$ & $\rm pH_2  $ & $\longrightarrow$  & $\rm pH_3^+$ & $\rm oH_2  $ & & 1.42(-09) & $-0.37$ & $177.24$ \\ 
$\rm pH_3^+$ & $\rm pH_2  $ & $\longrightarrow$  & $\rm oH_3^+$ & $\rm oH_2  $ & & 8.14(-10) & $-0.74$ & $227.61$ \\ 
$\rm pH_3^+$ & $\rm oH_2  $ & $\longrightarrow$  & $\rm pH_3^+$ & $\rm pH_2  $ & & 1.93(-10) & $-0.18$ & $1.03$ \\ 
$\rm pH_3^+$ & $\rm oH_2  $ & $\longrightarrow$  & $\rm oH_3^+$ & $\rm pH_2  $ & & 1.75(-10) & $-0.27$ & $1.66$ \\ 
$\rm pH_3^+$ & $\rm oH_2  $ & $\longrightarrow$  & $\rm oH_3^+$ & $\rm oH_2  $ & & 5.89(-10) & $-0.17$ & $35.67$ \\ 
$\rm pH_3^+$ & $\rm HD    $ & $\longrightarrow$  & $\rm oH_3^+$ & $\rm HD    $ & & 1.79(-10) & $0.59$ & $23.32$ \\ 
$\rm pH_3^+$ & $\rm HD    $ & $\longrightarrow$  & $\rm pH_2D^+$ & $\rm pH_2  $ & & 1.17(-10) & $-0.39$ & $2.70$ \\ 
$\rm pH_3^+$ & $\rm HD    $ & $\longrightarrow$  & $\rm pH_2D^+$ & $\rm oH_2  $ & & 2.51(-10) & $-0.27$ & $3.27$ \\
\end{longtable}

\end{document}